\documentclass[journal,twocolumn]{IEEEtran}
\usepackage{}
\usepackage{graphicx}
\usepackage{epstopdf}
\usepackage{amssymb}
\usepackage{amsfonts}
\usepackage{amsmath}
\usepackage{algorithm}
\usepackage{algorithmic}
\usepackage{subeqnarray}
\usepackage{cases}

\usepackage{subfigure,amsmath,amssymb,cite}
\usepackage{stfloats}

\usepackage{float}

\ifCLASSINFOpdf
\else
\fi
\begin{document}

\title{Beamforming Design for IRS-aided Decode-and-Forward Relay Wireless Network}

\author{Xuehui~Wang,~Feng Shu,~\emph{Member},~\emph{IEEE},~Weiping Shi,~Xiaopeng Liang,~Rongen Dong, Jun Li, ~\emph{Senior Member},~\emph{IEEE},~and~Jiangzhou~Wang,~\emph{Fellow},~\emph{IEEE}

\thanks{This work was supported in part by the National Natural Science Foundation of China (Nos. 62071234 and 61771244), and the Scientific Research Fund Project  of Hainan University under Grant KYQD(ZR)-21008 and KYQD(ZR)-21007).  \emph{(Corresponding authors: Feng Shu and Xiaopeng Liang)}.}

\thanks{Xuehui~Wang,~Feng Shu,~Xiaopeng Liang, and~Rongen Dong are  with the School of Information and Communication Engineering, Hainan University,~Haikou,~570228, China.}

\thanks{Weiping Shi,~Jun Li and Feng Shu are with the School of Electronic and Optical Engineering, Nanjing University of Science and Technology, 210094, China.}

\thanks{Jiangzhou Wang is with the School of Engineering, University of Kent, Canterbury CT2 7NT, U.K. Email: (e-mail: j.z.wang@kent.ac.uk).}

}

\maketitle

\begin{abstract}

As a low-cost and low-power-consumption passive reflector, intelligent reflecting surface (IRS) can make a significant rate improvement by building a programmable wireless environment. To improve the rate performance and coverage range of wireless networks, an IRS-aided decode-and-forward (DF) relay network is proposed with multiple antennas at relay station (RS).  To achieve a high rate, an alternately iterative structure (AIS) of maximizing receive power (Max-RP) at RS is proposed to jointly optimize the beamforming vectors at RS  and phase shifts at IRS.  Considering its high-complexity, two low-complexity Max-RP schemes of null-space projection (NSP) plus  maximum ratio combining (MRC)  and  IRS element selection (IRSES) plus MRC are presented to reduce  this complexity, respectively. For the former, NSP is used to separate the reflected signal from IRS and the direct transmitted signal from source and MRC is adopted to combine the two signals at RS. For the latter, the basic concept of IRSES is  as follows: IRS is partitioned into $M$ subsets of elements  and adjusting the phases of all elements per subset make all reflected signals and the direct signal from source phase alignment (PA) at the corresponding antenna of relay.  Simulation results show that the proposed three methods perform much better than the existing network with single-antenna relay  in terms of rate performance. In particular, a 85\%  rate gain over existing scheme is achieved in the high signal-to-noise ratio region. Moreover, it is verified that the positions of RS and IRS have a substantial impact on rate performance, and there exists an optimal positions of RS and IRS.

\end{abstract}



\section{Introduction}
With the explosive growth of communication device nodes in wireless sensor network (WSN), there are more stringent requirements in terms of energy efficiency and extended coverage \cite{8680631}. Relay has a strong ability to process signal for extended communication coverage and improved rate performance \cite{5502396}, \cite{4927440}. However, the conventional relay is an active device, which is costly and requires additional high energy consumption to process the signal \cite{DBLP:journals/corr/abs-2101-12091}, it is crucial to develop a energy-efficient and green communication network.

Compared with the conventional relay, since intelligent reflecting surface (IRS) does not require any radio frequency chain and baseband circuit and is made up of low-cost and passive reflecting units, its lower-energy-consumption and passive property is very attractive \cite{8910627},\cite{DBLP:journals/corr/abs-2108-06660}. Thus, IRS can be regarded as a green reflect-to-forward relay. It is anticipated that IRS will be potentially applied to the diverse future wireless networks such as WiFi 7, mobile communications  like sixth generation (6G), space communication, marine communication, and emergency communication \cite{2021Reconfigurable}.  For example, for mobile communications, IRS will be used to significantly enhance the coverage of blind areas, especially cell edges.  Due to the ability of intelligently adjusting the amplitude and phase shift of the incident signal by a smart controller \cite{9146177}, IRS may form helpful controlled multipaths, ingeniously improve the propagation environment and provide new degrees of freedom, which can be used to enhance the wireless network performance. Recently, IRS has attracted much attention from both academia and industry  \cite{8796365}. IRS may be adopted to aid many wireless communication directions as follows: directional modulation \cite{9530396}, \cite{9232092}, spatial modulation \cite{DBLP:journals/corr/abs-2106-03616,9481970}, multicast transmission \cite{9384498,2020Intelligent}, covert wireless communication \cite{9496108,9108996},  unmanned aerial vehicle (UAV) communication \cite{9367288,9293155},  and simultaneous wireless information and power transfer (SWIPT) \cite{2019Weighted,9288742}.

A combination of a relay and an IRS  was shown to improve energy efficiency \cite{9473608}, spectral efficiency \cite{9473487} and rate performance \cite{DBLP:journals/corr/abs-2012-12329,DBLP:journals/corr/abs-2006-06644,2020A}.  In \cite{9473608}, the authors proposed a multi-user mobile communication network model with the help of a relay and an IRS, where two sub-optimal methods based on alternating optimization (AO), called singular value decomposition (SVD) plus uplink-downlink duality, and SVD plus zero-forcing methods were presented to implement an enhanced performance from base station (BS) to multi-users in terms of energy efficiency. To increase spectral efficiency,  in \cite{9473487} a novel network was proposed, where some IRS elements act as active relays to amplify incident signals, and the remaining elements reflect signals. A spectral efficiency maximization problem was formulated and solved by the alternating optimization method.
Aiming at improving rate performance,  the authors proposed two kinds of IRS plus decode-and-forward (DF) relay \cite{DBLP:journals/corr/abs-2012-12329}: distributed and centralized. Subsequently, a sequential optimization algorithm was presented to address the power allocation and  optimization of IRS phases, and it was demonstrated that IRS and DF relay can work in a synergistic manner to enhance the achievable rate. In \cite{DBLP:journals/corr/abs-2006-06644}, a network consisting of two side-by-side intelligent surfaces connected via a full-duplex relay was proposed to achieve the promising rate gains with much smaller number of reflecting elements. Moreover, a hybrid network consisting of an IRS and a single-antenna DF relay was proposed to save a massive IRS elements compared with only IRS in \cite{2020A}. In \cite{9296320}, a communication system with multiple antennas at both transmitter and receiver is considered.

From the above literature, it can be concluded that IRS has the advantages of low cost and low energy consumption, and relay has a strong signal processing ability such as amplify-and-forward (AF) and DF. It is interesting to combine relay and IRS while keeping the advantages of relay and IRS. The hybrid network combines both advantages of IRS and relay to strike a good balance among circuit cost, energy efficiency and rate performance.

The application scenarios in the above literature are common, a hybrid network combining IRS and multi-antenna relay is proposed to employ to WSN scenario, which consists of a data collecting center and many other non-center nodes. The data collecting center collects all data from other nodes, then send them to internet with the help of relay and IRS. Furthermore, when data collecting center-relay or relay-destination direct link has shadow fading, or in extreme cases is completely blocked. The proposed hybrid network is still appropriate for the practical scenario, where an IRS is placed on a high big building  and can see source, relay and destination. In other words, the signal from source can still be transmitted to relay and destination with the help of the IRS. This means that adding IRS can create new reflective paths among source, relay and destination.

In this case, in order to improve the data rate of the proposed hybrid network or dramatically extend its coverage range, three efficient beamforming methods are proposed to achieve this goal. Our main contributions are summarized as follows:

\begin{enumerate}

\item
To make a dramatic rate improvement, an IRS-aided multi-antenna relay network model is proposed. We focus on the design of beamforming in the first time slot due to the fact that the system model can be converted into a typical IRS-aided three-point model. A maximizing receive power (Max-RP) using alternately iterative structure (AIS) is proposed to  maximize RP  by alternately  optimizing the beamforming vector at relay station (RS) and phase shifts at IRS.  Due to the rule of Max-RP,  the closed-form expressions of the beamforming vector at RS and phase shifts at IRS are derived.  From the simulation results, the proposed  Max-RP method using AIS can harvest up to 86\% rate gain over existing network of an IRS plus a single-antenna  relay  in \cite{2020A} in the high signal-to-noise ratio (SNR) region. The total computational complexity is $\mathcal{O}\{L_2({N^4}+8M{N^3}+5{N^3}+24M{N^2}-2{N^2})\}$, and the highest order is $N^4$ float-point operations (FLOPs), which is high.

\item

To reduce the high computational complexity of  the above method, a low-complexity null-space projection (NSP)-based Max-RP plus maximum ratio combining (MRC) is proposed in the first time slot. The two independent receive beamforming vectors are used at RS. They utilize NSP  to separate the signal from source and the reflected signal from IRS and maximize the receive power of the corresponding signal. Finally, MRC is adopted to combine the two signals to improve the SNR  of receive signal at RS. Our simulation results show that  the rate of the  proposed NSP-based Max-RP plus MRC method is 84\% higher than that of the existing system with an IRS and a single-antenna  relay  in \cite{2020A}  in the high SNR region. Its computational complexity is $\mathcal{O}\{N^3+M^3+{M^2}N+M{N^2}\}$, which is much lower than the Max-RP method using AIS.

\item
To further reduce the computational complexity, the IRS element selection (IRSES)-based Max-RP  plus MRC method is proposed. Here, in the first time slot, according to the number $M$ of RS antennas, IRS elements are divided into $M$ subsets with each subset having the same number of elements.  Each subset is mapped into one antenna at RS. In other words, there is one-to-one mapping relationship between subsets at IRS and antennas at RS. Adjusting the phases of all elements per subset aligns the phases of all reflected signals and the direct signal from source at the corresponding antenna of relay. Finally, MRC is adopted to combine all received signals at RS. From the simulation results, the proposed Max-RP based on IRSES plus MRC method achieves a substantial rate improvement over existing system in \cite{2020A}. The complexity is $\mathcal{O}\{15MK+8M+10K+L_5(18MN+2M+3N)\}$, which is extremely lower than the above proposed two methods.

\end{enumerate}

The remainder of this paper is organized as follows. In Section II, we describe an IRS-aided DF relay network. In Section III, we demonstrate three methods for a better rate performance of the proposed network. We present our simulation results in Section IV, and draw conclusions in Section V.

\emph{Notation}: Scalars, vectors and matrices are respectively represented by letters of lower case, bold lower case, and bold upper case. $(\cdot)^*$, $(\cdot)^H$, $(\cdot)^\dagger$ stand for matrix conjugate, conjugate transpose, and Moore-Penrose pseudo inverse, respectively.
$\mathbb{E}\{\cdot\}$, $\|\cdot\|$, $\text{tr}(\cdot)$, and $\text{arg}(\cdot)$ denote expectation operation, 2-norm, the trace of a matrix, and the phase of a complex number, respectively.
The sign $\textbf{I}_{N}$ is the $N\times N$ identity matrix.

\section{System Model}
\begin{figure*}[htbp]
\makebox[\textwidth][c]{\includegraphics[width=0.9\textwidth]{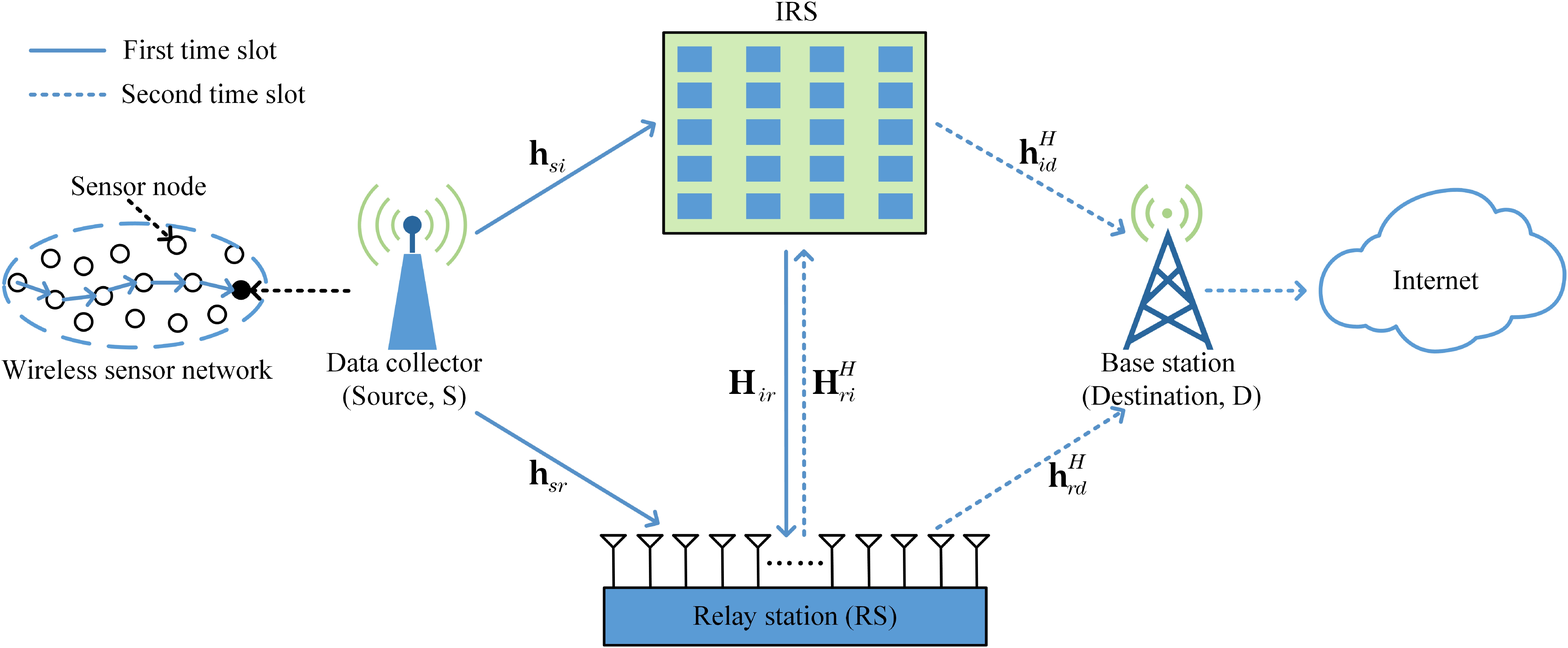}}
\caption{System model for an IRS-aided DF relay network.}  \label{1System-model.eps}
\end{figure*}
As shown in Fig. 1, the IRS-aided DF relay network consists of a half-duplex DF relay station (RS) with $M$ transmit antennas, an IRS with $N$ reflecting elements, a data collector (S) and a base station (D) equipped with a single antenna each. As a data collecting center node, S collects all data from other sensor nodes, and transmits them towards D with the help of IRS and RS. Here, the network is operated in a time division half-duplex scenario. The distance from S to D are assumed so far that there is no direct link between them \cite{2020A}. Due to path loss, the power of signals  reflected by the IRS twice or more are such weak that they can be ignored. Moreover, all channel state informations (CSIs) are assumed to be perfectly known to S, RS, IRS and D. In the first time slot, the received signal at RS is written by
\begin{equation}
\textbf{y}_r=\sqrt {P_s}( \textbf{h}_{sr}+\textbf{H}_{ir}{\boldsymbol\Theta}_1\textbf{h}_{si})s+\textbf{n}_r,
\end{equation}
where $s$ and $ {P_s}$ are the transmit signal and power from S, respectively. $\mathbb{E}\{s^Hs\}=1$. Without loss
of generality, we assume a Rayleigh fading environment. Let $\textbf{h}_{sr}$$\in \mathbb C^{M \times 1}$, $\textbf{H}_{ir}\in \mathbb C^{M \times N}$ and $\textbf{h}_{si}\in \mathbb C^{N \times 1}$ represent the channels from S to RS, from IRS to RS, and from S to IRS. ${\boldsymbol\Theta}_1$ is the diagonal reflection-coefficient matrix of IRS, which is denoted as $ {\boldsymbol\Theta}_1$ = $\text{diag}( e^{j\theta _{11}}, \cdots,e^{j\theta _{1N}} )$, ${\boldsymbol\theta}_1 = {[e^{j\theta _{11}}}, \cdots ,{e^{j\theta _{1N}}]^T}$, $\theta _{1i}\in ( 0 ,{2\pi } ]$ is the phase shift of the $i$th element. $\textbf{n}_r\in \mathbb C^{M \times 1}$ is the additive white Gaussian noise (AWGN) with distribution $\textbf{n}_r\sim \mathcal{CN}( 0,\sigma _r^2{\bf I}_{N_r} )$.
Then the received signal at RS after receive beamforming can be expressed as
\begin{equation}
y_r = \sqrt {P_s}\textbf{u}_r^H( \textbf{h}_{sr}+\textbf{H}_{ir}{\boldsymbol\Theta}_1\textbf{h}_{si})s + \textbf{u}_r^H\textbf{n}_r,
\end{equation}
where $\textbf{u}_r^H\in \mathbb C^{1\times M}$ is the receive beamforming vector, $\|\textbf{u}_r^H\|^2=1$. The achievable rate at RS is given by
\begin{equation}\label{R_r}
R_r=\log_2\left(1+\frac{ P_R}{\sigma _r^2}\right),
\end{equation}
where
\begin{equation}
P_R=P_s\text{tr}\left\{{{{ \textbf{u}_r^H}(\textbf{h}_{sr}+\textbf{H}_{ir}{\boldsymbol\Theta}_1\textbf{h}_{si})(\textbf{h}_{sr}+\textbf{H}_{ir}{\boldsymbol\Theta}_1\textbf{h}_{si})^H\textbf{u}_r}}\right\}.
\end{equation}

In the second time slot, it is assumed that the original signal $s$ can be correctly decoded by RS. The transmit signal from RS is $\textbf{x}_r=\textbf{u}_ts$,
where $\textbf{u}_t\in \mathbb{C}^{M \times 1}$ is the transmit beamforming vector, $\|\textbf{u}_t\|^2=1$. The received signal at D is as follows
\begin{equation}
y_{d} = \sqrt {P_r}( \textbf{h}_{rd}^H+\textbf{h}_{id}^H{\boldsymbol\Theta}_2\textbf{H}_{ri}^H)\textbf{u}_ts + n_{d},
\end{equation}
where ${P_r}$ is the transmit power from RS, $\textbf{h}_{rd}^H \in {\mathbb{C}^{1 \times M}}$, $\textbf{h}_{id}^H \in {\mathbb{C}^{1 \times N}}$ and $\textbf{H}_{ri}^H \in {\mathbb{C}^{N\times M}}$ are the channels from RS to D, from IRS to D, and from RS to IRS, respectively. The diagonal reflection-coefficient matrix of IRS is represented as ${\boldsymbol\Theta}_2 = \text{diag}\left( e^{j{\theta _{21}}}, \cdots ,e^{j{\theta _{2N}}} \right)$, ${\boldsymbol\theta}_2 = {[e^{j\theta _{21}}}, \cdots ,{e^{j\theta _{2N}}]^T}$, ${\theta _{2i}}\in \left( 0 \right.,\left. {2\pi } \right]$ is the phase shift of the $i$th element. ${n_{d}}$ is the AWGN with distribution $n_{d} \sim \mathcal{CN}\left( 0,{\sigma_{d}^2} \right)$.
The achievable rate at D can be expressed as
\begin{equation}
R_d=\log_2\left(1 + \frac{P_D }{\sigma _d^2}\right),
\end{equation}
where
\begin{equation}
P_D=P_r\text{tr}\left\{ \textbf{u}_t^H(\textbf{h}_{rd}^H+\textbf{h}_{id}^H{\boldsymbol\Theta}_2\textbf{H}_{ri}^H)^H(\textbf{h}_{rd}^H + \textbf{h}_{id}^H{\boldsymbol\Theta}_2\textbf{H}_{ri}^H)\textbf{u}_t\right\}.
\end{equation}

The achievable rate of the proposed system is defined as follows
\begin{equation}\label{system rate}
R_s=\frac{1}{2}\text{min}\{R_r,R_d  \}.
\end{equation}

\section{Proposed Three High-Performance Beamforming Schemes}

In this section, according to (\ref{R_r}),  maximizing rate is equivalent to Max-RP due to the fact that the log function is a monotone increasing function of RP. In the first time slot, three Max-RP methods: AIS-based Max-RP, NSP-based Max-RP plus MRC, and IRSES-based Max-RP plus MRC, are proposed to optimize the phase shift vector ${\boldsymbol\theta}_1$ of IRS and the receive beamforming vector $\textbf{u}_r$ of RS. In the second time slot, RS, IRS and destination form a typical three-point IRS-aided network, where an alternating iteration in \cite{8811733} are adopted to design the transmit beamforming vector at RS and adjust the phases of IRS.

\subsection{Proposed AIS-based Max-RP Method}

For the first time slot, the corresponding optimization problem can be casted as
\begin{align}\label{p1}
&\max \limits_{{\boldsymbol\Theta}_1,\textbf{u}_r^H}~~~~P_s\text{tr}\left\{{{{
\textbf{u}_r^H}(\textbf{h}_{sr}+\textbf{H}_{ir}{\boldsymbol\Theta}_1\textbf{h}_{si})(\textbf{h}_{sr}+\textbf{H}_{ir}{\boldsymbol\Theta}_1\textbf{h}_{si})^H\textbf{u}_r}}\right\}\nonumber\\
&~~\text{s.t.}~~~~~~ \|\textbf{u}_r^H\|^2=1,~|\boldsymbol{\theta}_1(i)|=1,~\forall i = 1, \cdots ,N.
\end{align}

Let us define  $\textbf{H}_{si}= \text{diag}\{\textbf{h}_{si}\}$, we have ${\boldsymbol\Theta}_1\textbf{h}_{si}=\textbf{H}_{si}{\boldsymbol\theta }_1$. Firstly, by fixing $\textbf{u}_r^H$, (\ref{p1}) is reduced to
\begin{align}\label{p2}
&\max \limits_{{\boldsymbol\theta } _1}~~~P_s\text{tr}({\boldsymbol\theta }_1^H\textbf{H}_{si}^H\textbf{H}_{ir}^H\textbf{u}_{r}\textbf{u}_{r}^H\textbf{H}_{ir}\textbf{H}_{si}{\boldsymbol\theta }_1)+\nonumber\\
&~~~~~~~~~P_s\text{tr}({\boldsymbol\theta }_1^H\textbf{H}_{si}^H\textbf{H}_{ir}^H\textbf{u}_r\textbf{u}_r^H\textbf{h}_{sr})+
P_s\text{tr}({\boldsymbol\theta }_1\textbf{h}_{sr}^H\textbf{u}_{r}\textbf{u}_r^H\textbf{H}_{ir}\textbf{H}_{si})\nonumber\\
&~\text{s.t.}~~~~~{\boldsymbol\theta}_{1}^H{\boldsymbol\theta}_{1}= N.
\end{align}
The Lagrangian function associated with the above optimization (\ref{p2}) is defined as
\begin{align}\label{lagrangian function}
&L({\boldsymbol\theta }_1,\lambda)=P_s\text{tr}({\boldsymbol\theta}_1^H\textbf{H}_{si}^H\textbf{H}_{ir}^H\textbf{u}_{r}\textbf{u}_{r}^H\textbf{H}_{ir}\textbf{H}_{si}{\boldsymbol\theta }_1)+\nonumber\\
&~~~~P_s\text{tr}({\boldsymbol\theta }_1^H\textbf{H}_{si}^H\textbf{H}_{ir}^H\textbf{u}_r\textbf{u}_r^H\textbf{h}_{sr})+
P_s\text{tr}({\boldsymbol\theta }_1\textbf{h}_{sr}^H\textbf{u}_{r}\textbf{u}_r^H\textbf{H}_{ir}\textbf{H}_{si})+\nonumber\\
&~~~~~~~~~~~~~~~~~~~~~~~~~~~~~~~~~~~~~~~~~~~~\lambda( {{\boldsymbol\theta}_{1}^H{\boldsymbol\theta}_{1}}- N),
\end{align}
where $\lambda$ is the Lagrange multiplier, if (${\boldsymbol\theta }_1$, $\lambda$) is the optimal solution to the above equation, the partial derivative of the Lagrangian function with respect to ${\boldsymbol\theta}_1$ should be set to be zero as follows
\begin{align}
&\frac{\partial L({\boldsymbol\theta}_{1},\lambda)}{\partial {\boldsymbol\theta}_{1}}=P_s\textbf{H}_{si}^H\textbf{H}_{ir}^H\textbf{u}_{r} \textbf{u}_{r}^H\textbf{H}_{ir}\textbf{H}_{si}{\boldsymbol\theta}_1+\nonumber\\
&~~~~~~~~~~~~~~~~~~~~~~~~~P_s\textbf{H}_{si}^H\textbf{H}_{ir}^H\textbf{u}_{r}\textbf{u}_{r}^H\textbf{h}_{sr}+\lambda{\boldsymbol\theta}_1\nonumber\\
&~~~~~~~~~~~~~=0,
\end{align}
which gives the solution
\begin{align}\label{T1}
&\theta _{1i}=-\text{arg}\{(\lambda
{P_s}^{-1}\textbf{I}_{N}+\textbf{H}_{si}^H\textbf{H}_{ir}^H\textbf{u}_{r}\textbf{u}_{r}^H\textbf{H}_{ir}\textbf{H}_{si})^{\dagger}\cdot\nonumber\\
&~~~~~~~~~~~~~~~~~~~~~~~~~~~~~~~~~~~~(\textbf{H}_{si}^H\textbf{H}_{ir}^H\textbf{u}_{r}\textbf{u}_{r}^H\textbf{h}_{sr})\}_i.
\end{align}
Inserting (\ref{T1}) back into the restriction of (\ref{p2}), we have an equation associated with Lagrange multiplier as follows
\begin{equation}\label{det1}
\det\left( \lambda{P_s}^{-1}\textbf{I}_{N}+\textbf{H}_{si}^H\textbf{H}_{ir}^H\textbf{u}_{r}\textbf{u}_{r}^H\textbf{H}_{ir}\textbf{H}_{si}\right)=b,
\end{equation}
where
\begin{equation}
b=\left(\det(N^{-1}\textbf{H}_{si}^H\textbf{H}_{ir}^H\textbf{u}_{r}\textbf{u}_{r}^H\textbf{h}_{sr}\textbf{h}_{sr}^H\textbf{u}_{r}\textbf{u}_{r}^H\textbf{H}_{ir}\textbf{H}_{si}) \right)^\frac{1}{2}.
\end{equation}
Further, in accordance with the rank inequality: $\text{rank}~\textbf{A} \leq \min \{m, n\}$, where $\textbf{A}\in \mathbb C^{m \times n}$, we can get
\begin{equation}\label{rank}
\text{rank}(\textbf{H}_{si}^H\textbf{H}_{ir}^H\textbf{u}_{r}\textbf{u}_{r}^H\textbf{h}_{sr}\textbf{h}_{sr}^H\textbf{u}_{r}\textbf{u}_{r}^H\textbf{H}_{ir}\textbf{H}_{si})\leq \text{rank}(\textbf{u}_{r}^H)=1.
\end{equation}
Therefore, $b=0$.
Then we have the eigen decomposition as follows
\begin{equation}\label{eigen decomposition}
\textbf{H}_{si}^H\textbf{H}_{ir}^H\textbf{u}_{r}\textbf{u}_{r}^H\textbf{H}_{ir}\textbf{H}_{si}=\textbf{U}\boldsymbol\Sigma \textbf{U}^H,
\end{equation}
where $\textbf{U}$ is a unitary matrix composed of eigenvectors and $\textbf{U}\textbf{U}^H$=$\textbf{U}^H\textbf{U}$=$\textbf{I}_{N}$.
$\boldsymbol\Sigma$ is a diagonal matrix composed of eigenvalues.
Substituting the above equation in (\ref{det1}) yields the following simplified equation
\begin{align}\label{det2}
&\det\left\{\textbf{U}\left( \lambda{P_s}^{-1}\textbf{I}_{N}+\boldsymbol\Sigma \right)\textbf{U}^H\right\}=\det\left\{\textbf{U}\textbf{U}^H\left( \lambda{P_s}^{-1}\textbf{I}_{N}+\boldsymbol\Sigma \right)\right\}\nonumber\\
&~~~~~~~~~~~~~~~~~~~~~~~~~~~~~~~~~~~=\det\left( \lambda{P_s}^{-1}\textbf{I}_{N}+\boldsymbol\Sigma \right)\nonumber\\
&~~~~~~~~~~~~~~~~~~~~~~~~~~~~~~~~~~~=0.
\end{align}
It can be observed that $\text{rank}(\textbf{H}_{si}^H\textbf{H}_{ir}^H\textbf{u}_{r}\textbf{u}_{r}^H\textbf{H}_{ir}\textbf{H}_{si}) = \text{rank}(\textbf{H}_{si}^H\textbf{H}_{ir}^H\textbf{u}_{r}) = \text{rank}(\textbf{u}_{r}^H\textbf{H}_{ir}\textbf{H}_{si}) = 1$, so there is a nonzero eigenvalue and $N-1$ zero eigenvalues in $\textbf{H}_{si}^H\textbf{H}_{ir}^H\textbf{u}_{r}\textbf{u}_{r}^H\textbf{H}_{ir}\textbf{H}_{si}$. Then we construct the following equation
\begin{equation}\label{nonzero eigenvalue}
\textbf{H}_{si}^H\textbf{H}_{ir}^H\textbf{u}_{r}\textbf{u}_{r}^H\textbf{H}_{ir}\textbf{H}_{si}(\textbf{H}_{si}^H\textbf{H}_{ir}^H\textbf{u}_{r})=\|\textbf{u}_{r}^H\textbf{H}_{ir}\textbf{H}_{si}\|^2\textbf{H}_{si}^H\textbf{H}_{ir}^H\textbf{u}_{r},
\end{equation}
where $\|\textbf{u}_{r}^H\textbf{H}_{ir}\textbf{H}_{si}\|^2$ is the only nonzero eigenvalue (ie. the largest eigenvalue), so we have
\begin{equation}\label{Sigma}
\boldsymbol\Sigma =\left[
\begin{array}{cccc}
\|\textbf{u}_{r}^H\textbf{H}_{ir}\textbf{H}_{si}\|^2 & 0 & \cdots & 0\\
0 & 0 & \cdots & 0\\
\vdots & \vdots & \ddots & \vdots\\
0 & 0 & \cdots & 0
\end{array}
\right]_{N \times N}.
\end{equation}
Substituting (\ref{Sigma}) into (\ref{det2}) yields
\begin{equation}
\left|\begin{array}{cccc}
    \lambda+\|\textbf{u}_{r}^H\textbf{H}_{ir}\textbf{H}_{si}\|^2{P_s} &    0   & \cdots & 0\\
    0 &    \lambda   & \cdots & 0\\
    \vdots & \vdots & \ddots & \vdots\\
0 & 0 & \cdots & \lambda
\end{array}\right|_{N \times N}=0,
\end{equation}
which is expanded as follows
\begin{equation}\label{lambda1}
\lambda^N+\|\textbf{u}_{r}^H\textbf{H}_{ir}\textbf{H}_{si}\|^2{P_s}\lambda^{N-1}=0,
\end{equation}
which forms the set of candidate solutions
\begin{equation}\label{lambda2}
\lambda\in\{0, -\|\textbf{u}_{r}^H\textbf{H}_{ir}\textbf{H}_{si}\|^2{P_s} \}.
\end{equation}
Plugging (\ref{lambda2}) into (\ref{T1}), we find that $\lambda= -\|\textbf{u}_{r}^H\textbf{H}_{ir}\textbf{H}_{si}\|^2{P_s}$ is not suitable for (\ref{T1}). Inserting $\lambda=0$ back into (\ref{T1}) yields the following solution
\begin{equation}
\theta _{1i}=-\text{arg}\{(\textbf{H}_{si}^H\textbf{H}_{ir}^H\textbf{u}_{r}\textbf{u}_{r}^H\textbf{H}_{ir}\textbf{H}_{si})^{\dagger}
(\textbf{H}_{si}^H\textbf{H}_{ir}^H\textbf{u}_{r}\textbf{u}_{r}^H\textbf{h}_{sr})\}_i.
\end{equation}

And then by fixing $\boldsymbol{\Theta}_{1}$, the optimization problem in (\ref{p1}) can be converted into
\begin{align}\label{p3}
&\max \limits_{\textbf{u}_r^H}~~~P_s\text{tr}({{{\textbf{u}_r^H}\textbf{B}\textbf{u}_r}})\nonumber\\
&\ \text{s.t.}~~~~~\|\textbf{u}_r^H\|^2=1,
\end{align}
where \begin{equation}
\textbf{B}=(\textbf{h}_{sr}+\textbf{H}_{ir}{\boldsymbol\Theta}_1\textbf{h}_{si})(\textbf{h}_{sr}+\textbf{H}_{ir}{\boldsymbol\Theta}_1\textbf{h}_{si})^H,
\end{equation}
is a Hermitian symmetric matrix. Similar to (\ref{lagrangian function}), we have
\begin{equation}
L(\textbf{u}_r,\mu)=P_s\text{tr}(\textbf{u}_r^H\textbf{B}\textbf{u}_r)+\mu(\|\textbf{u}_r^H\|^2-1),
\end{equation}
where $\mu$ denotes the Lagrange multiplier. Setting the partial derivative of the Lagrange function $L(\textbf{u}_r,\mu)$ with respect to $\textbf{u}_r$ equals 0, we have
\begin{equation}\label{eigenvector}
P_s\textbf{B}\textbf{u}_r+\mu\textbf{u}_r=0.
\end{equation}
Similar to (\ref{rank}), rank(\textbf{B})$=1$, so there is a nonzero eigenvalue in $\textbf{B}$. It indicates that $\textbf{u}_r$ is the eigenvector corresponding to the nonzero eigenvalue of the matrix $\textbf{B}$. Similar to (\ref{nonzero eigenvalue}), we obtain the solution
\begin{equation}
\textbf{u}_r=\frac{\textbf{h}_{sr}+\textbf{H}_{ir}{\boldsymbol\Theta}_1\textbf{h}_{si}}{\| \textbf{h}_{sr}+\textbf{H}_{ir}{\boldsymbol\Theta}_1\textbf{h}_{si} \|}.
\end{equation}

It can be observed that when one of the  optimization variables ${\boldsymbol\Theta}_1$ and $\textbf{u}_r^H$ is fixed, the objective function in (9) is convex with respect to the other variable, in each iteration the objective function value monotonically increases with a finite upper-bound. Therefore, it can be guaranteed that the alternately iterative (AI) algorithm is convergent. AI  between $\theta _{1i}$ and $\textbf{u}_r$ are performed until the condition of convergence is reached. When $\theta _{1i}$ and $\textbf{u}_r$ are optimal, the $R_r$ in (\ref{R_r}) is maximized. The proposed alternate iteration algorithm for obtaining an optimal solution $R_r$ is summarized in Algorithm 1.
\begin{table}[h]\normalsize
\renewcommand{\arraystretch}{1}
\centering
\begin{tabular}{p{240pt}}
\hline
$\bf{Algorithm~1}$  Proposed alternate iteration algorithm\\
\hline
1.\ Initialize the receive beamforming $\textbf{u}_r^1$, set the convergence error $\varepsilon$ and the iteration number $t$ = 1. \\
2. \bf{repeat}  \\
3.\ \ \  Solve (\ref{p2}) for a given $\textbf{u}_r^t$, denote the optimal solution as ${\boldsymbol\theta}_{1}^t$ and $\theta _{1i}^t$.\\
4.\ \ \  Solve (\ref{p3}) for a given ${\boldsymbol\theta}_{1}^t$,  denote the optimal solution as $\textbf{u}_r^{t + 1}$.\\
5.\ \ \  Update $t$ = $t$+1.\\
6. \bf{until} \\
\ \ \ \ \    $\left| {R_r^{t+1} - R_r^t} \right|\le \varepsilon$.\\
\hline
\end{tabular}
\end{table}

According to (\ref{system rate}), we make a comparison between $R_r$ and $R_d$, and multiply the minimum value by 1/2 to obtain the system rate $R_s$.
The total complexity of proposed AIS-based Max-RP method is obtained as follows
\begin{align}
&\mathcal{O}\{L_2({N^4}+8M{N^3}+5{N^3}+24M{N^2}-2{N^2})+(3L_1+\nonumber\\
&~~~~~18L_2)MN+(5L_1+2L_2)M+(4L_1+3L_2  )N\},
\end{align}
where $L_1$ and $L_2$ are the AI numbers in the first time slot and the second time slot, respectively. Obviously, the highest order of computational complexity is $N^4$ float-point operations (FLOPs), which is high. So a low-complexity NSP-based Max-RP plus MRC method is proposed to reduce the computational complexity in the next subsection.

\subsection{Proposed NSP-based Max-RP plus MRC Method}

Considering the two signals from IRS and source node overlap together and interfere with each other,  in order to explore the two-way diversity gain, two individual receiving beamforming vectors at RS are used to separate them  respectively and MRC is used to combine them to exploit the multiple diversities.  After beamforming, the received signals at RS are rewritten by
\begin{equation}\label{yrs}
y_{rs} = \sqrt {P_s}\textbf{u}_{rs}^H\textbf{h}_{sr}s+\sqrt {P_s}\textbf{u}_{rs}^H\textbf{H}_{ir}{\boldsymbol\Theta}_1\textbf{h}_{si}s + \textbf{u}_{rs}^H\textbf{n}_{r},
\end{equation}
\begin{equation}\label{yri}
y_{ri} = \sqrt {P_s}\textbf{u}_{ri}^H\textbf{h}_{sr}s+\sqrt {P_s}\textbf{u}_{ri}^H\textbf{H}_{ir}{\boldsymbol\Theta}_1\textbf{h}_{si}s + \textbf{u}_{ri}^H\textbf{n}_{r},
\end{equation}
where $\textbf{u}_{rs}^H\in \mathbb C^{1\times M}$ and $\textbf{u}_{ri}^H\in \mathbb C^{1\times M}$ are defined as the beamforming vectors for S and IRS, respectively, which extract the independent signals from S and IRS, respectively.
$\|\textbf{u}_{rs}^H\|^2=1$ and $\|\textbf{u}_{ri}^H\|^2=1$.
We assume that $\textbf{u}_{rs}^H$ and $\textbf{u}_{ri}^H$ are in the null-space of channel $\textbf{H}_{ir}$ and $\textbf{h}_{sr}$, respectively. It is obvious that
\begin{equation}
\textbf{u}_{rs}^H\textbf{H}_{ir}=\textbf{0}^T, \ \textbf{u}_{ri}^H\textbf{h}_{sr}=0,
\end{equation}
which mean that the signals from S and IRS do not interfere with each other at RS.
Correspondingly, (\ref{yrs}) and (\ref{yri}) are reduced to
\begin{equation}
y_{rs} = \sqrt {P_s}\textbf{u}_{rs}^H\textbf{h}_{sr}s + \textbf{u}_{rs}^H\textbf{n}_{r},
\end{equation}
\begin{equation}
y_{ri} = \sqrt {P_s}\textbf{u}_{ri}^H\textbf{H}_{ir}{\boldsymbol\Theta}_1\textbf{h}_{si}s + \textbf{u}_{ri}^H\textbf{n}_{r}.
\end{equation}

Maximizing $R_r$ is equal to maximizing the power of $y_{rs}$ and $y_{ri}$, respectively. The
corresponding optimization problem of the power of $y_{rs}$ is given by
\begin{align}\label{Eyrs}
&\max \limits_{\textbf{u}_{rs}^H}~~~~P_s\text{tr}(\textbf{u}_{rs}^H\textbf{h}_{sr}\textbf{h}_{sr}^H\textbf{u}_{rs})\nonumber\\
&~\text{s.t.}~~~~~~\|\textbf{u}_{rs}^H\|^2=1,~\textbf{u}_{rs}^H\textbf{H}_{ir}=\textbf{0}^T.
\end{align}
According to the second constraint of the above optimization problem, let us define $\textbf{u}_{rs}$ as follows
\begin{equation}
\textbf{u}_{rs}=(\textbf{I}_{M}-\textbf{H}_{ir}(\textbf{H}_{ir}^H\textbf{H}_{ir})^{\dagger}\textbf{H}_{ir}^H)\textbf{v}_{rs}=\textbf{P}\textbf{v}_{rs} ,
\end{equation}
where $\textbf{P}=\textbf{I}_{M}-\textbf{H}_{ir}(\textbf{H}_{ir}^H\textbf{H}_{ir})^{\dagger}\textbf{H}_{ir}^H$, $\textbf{v}_{rs}$ is a new variable and $\|\textbf{v}_{rs} \|=1$. Inserting the above equation back into (\ref{Eyrs}), we obtain the simplified optimization problem as follows
\begin{align}
&\max \limits_{\textbf{v}_{rs}^H}~~~~P_s\textbf{v}_{rs}^H\textbf{P}^H \textbf{h}_{sr}\textbf{h}_{sr}^H\textbf{P}\textbf{v}_{rs}\nonumber\\
&~\text{s.t.}~~~~~~\|\textbf{v}_{rs}\|^2=1.
\end{align}
Similar to (\ref{p3}), the solution $\textbf{v}_{rs}$ is given by
\begin{equation}
\textbf{v}_{rs}=\frac{\textbf{P}^H \textbf{h}_{sr}}{\| \textbf{P}^H \textbf{h}_{sr}   \|} ,
\end{equation}
which yields
\begin{equation}
\textbf{u}_{rs}=\frac{\textbf{P} \textbf{v}_{rs}}{\| \textbf{P} \textbf{v}_{rs} \|} =\frac{(\textbf{I}_{M}-\textbf{H}_{ir}(\textbf{H}_{ir}^H\textbf{H}_{ir})^{\dagger}\textbf{H}_{ir}^H)^2\textbf{h}_{sr}}{\|(\textbf{I}_{M}-\textbf{H}_{ir}(\textbf{H}_{ir}^H\textbf{H}_{ir})^{\dagger}\textbf{H}_{ir}^H)^2\textbf{h}_{sr}\| } .
\end{equation}

In the same manner, the optimization problem of the power of $y_{ri}$ is also written by
\begin{align}\label{Eyris}
&\max \limits_{\textbf{u}_{ri}^H,{\boldsymbol\theta }_1}~~~~P_s\text{tr}(\textbf{u}_{ri}^H\textbf{H}_{ir}\textbf{H}_{si}{\boldsymbol\theta }_1{\boldsymbol\theta }_1^H\textbf{H}_{si}^H\textbf{H}_{ir}^H\textbf{u}_{ri})\nonumber\\
&~\text{s.t.}~~~~~~\|\textbf{u}_{ri}^H\|^2=1,~\textbf{u}_{ri}^H\textbf{h}_{sr}=0,~{\boldsymbol\theta}_{1}^H{\boldsymbol\theta}_{1}= N.
\end{align}
For a given $\textbf{u}_{ri}$, the solution $\theta_{1i}$ can be expressed as
\begin{equation}
\theta_{1i}
=-\text{arg}\{\textbf{H}_{si}^H\textbf{H}_{ir}^H\textbf{u}_{ri}\}_i.
\end{equation}
For a given ${\boldsymbol\theta }_1$, the solution $\textbf{u}_{ri}$ can be represented as
\begin{equation}
\textbf{u}_{ri}=\frac{(\textbf{I}_{M}-\textbf{h}_{sr}(\textbf{h}_{sr}^H\textbf{h}_{sr})^{\dagger}\textbf{h}_{sr}^H)^2\textbf{H}_{ir}\textbf{H}_{si}{\boldsymbol\theta }_1}{\| (\textbf{I}_{M}-\textbf{h}_{sr}(\textbf{h}_{sr}^H\textbf{h}_{sr})^{\dagger}\textbf{h}_{sr}^H)^2\textbf{H}_{ir}\textbf{H}_{si}{\boldsymbol\theta }_1\| }.
\end{equation}
Similar to Algorithm 1, $\textbf{u}_{ri}$ and ${\boldsymbol\theta }_1 $ are performed via AI procedure to obtain the maximum power of $y_{ri}$.
Then the joint received signal at RS by applying MRC can be expressed as
\begin{align}
&y_{r} =\frac{(\textbf{u}_{rs}^H\textbf{h}_{sr})^Hy_{rs}+ (\textbf{u}_{ri}^H\textbf{H}_{ir}\textbf{H}_{si}\boldsymbol{\theta_1})^Hy_{ri}} {\|\textbf{u}_{rs}^H\textbf{h}_{sr}+\textbf{u}_{ri}^H\textbf{H}_{ir}\textbf{H}_{si}\boldsymbol{\theta_1}\|}\nonumber\\
&=\sqrt {P_s}\left(\frac{\textbf{h}_{sr}^H\textbf{u}_{rs}\textbf{u}_{rs}^H\textbf{h}_{sr}+ \boldsymbol\theta_1^{H}\textbf{H}_{si}^{H}\textbf{H}_{ir}^{H}\textbf{u}_{ri}\textbf{u}_{ri}^H\textbf{H}_{ir}\textbf{H}_{si}\boldsymbol{\theta_1} }{\|\textbf{u}_{rs}^H\textbf{h}_{sr}+\textbf{u}_{ri}^H\textbf{H}_{ir}\textbf{H}_{si}\boldsymbol{\theta_1}\|}\right)s+\nonumber\\
&~~~~~~~~~~\left(\frac{\textbf{h}_{sr}^H\textbf{u}_{rs}\textbf{u}_{rs}^H+\boldsymbol\theta_1^{H}\textbf{H}_{si}^{H}\textbf{H}_{ir}^{H}\textbf{u}_{ri}\textbf{u}_{ri}^H   } {\|\textbf{u}_{rs}^H\textbf{h}_{sr}+\textbf{u}_{ri}^H\textbf{H}_{ir}\textbf{H}_{si}\boldsymbol{\theta_1}\|}\right)\textbf{n}_{r}.
\end{align}
The signals from S and IRS received by RS are independent, thus the achievable rate at RS can be calculated as
\begin{equation}
R_r=\log_2\left(1+\frac{(\|\textbf{u}_{rs}^H\textbf{h}_{sr}\|^4+ \|\textbf{u}_{ri}^H\textbf{H}_{ir}\textbf{H}_{si}\boldsymbol{\theta_1}\|^4 )P_s }{\|\textbf{u}_{rs}^H\textbf{h}_{sr}+\textbf{u}_{ri}^H\textbf{H}_{ir}\textbf{H}_{si}\boldsymbol{\theta_1}\|^2\sigma _r^2}\right).
\end{equation}

Compare $R_r$ and $R_d$ to obtain the system rate $R_s$. The complexity of the proposed NSP-based Max-RP plus MRC method is written by
\begin{align}
&\mathcal{O}\{{N^3}+2(1+L_3){M^3}+2(1+L_3){M^2}N+\nonumber\\
&~~~~~~~~~~~(4+3L_3)M{N^2}+(4+3L_3){M^2}-L_3{N^2}-\nonumber\\
&~~~~~~~~~~~~~~~(1-5L_3-18L_4)MN-(1-4L_3-2L_4)M+\nonumber\\
&~~~~~~~~~~~~~~~~~~~(1+2L_3+3L_4)N\},
\end{align}
where $L_3$ and $L_4$ respectively denote the AI numbers in the first time slot and the second time slot. The highest order of complexity is $N^3$ FLOPs, which is lower than AIS-based Max-RP Method and still very high. To further reduce the computational complexity, an IRSES-based Max-RP plus MRC method is proposed in the following subsection.

\subsection{Proposed IRSES-based Max-RP plus MRC Method}

As shown in Fig. 2, according to the number $M$ of RS antennas, the $N$ elements in IRS are randomly divided into $M$ suarrays evenly, each of which is mapped into one antenna at RS with $\frac{N}{M}$ = $K$ elements. Fig. 3 is the block diagram for IRSES-based Max-RP plus MRC method at RS, the details are described as below.
\begin{figure}[h]
\centering
\includegraphics[width=0.48\textwidth,height=0.27\textheight]{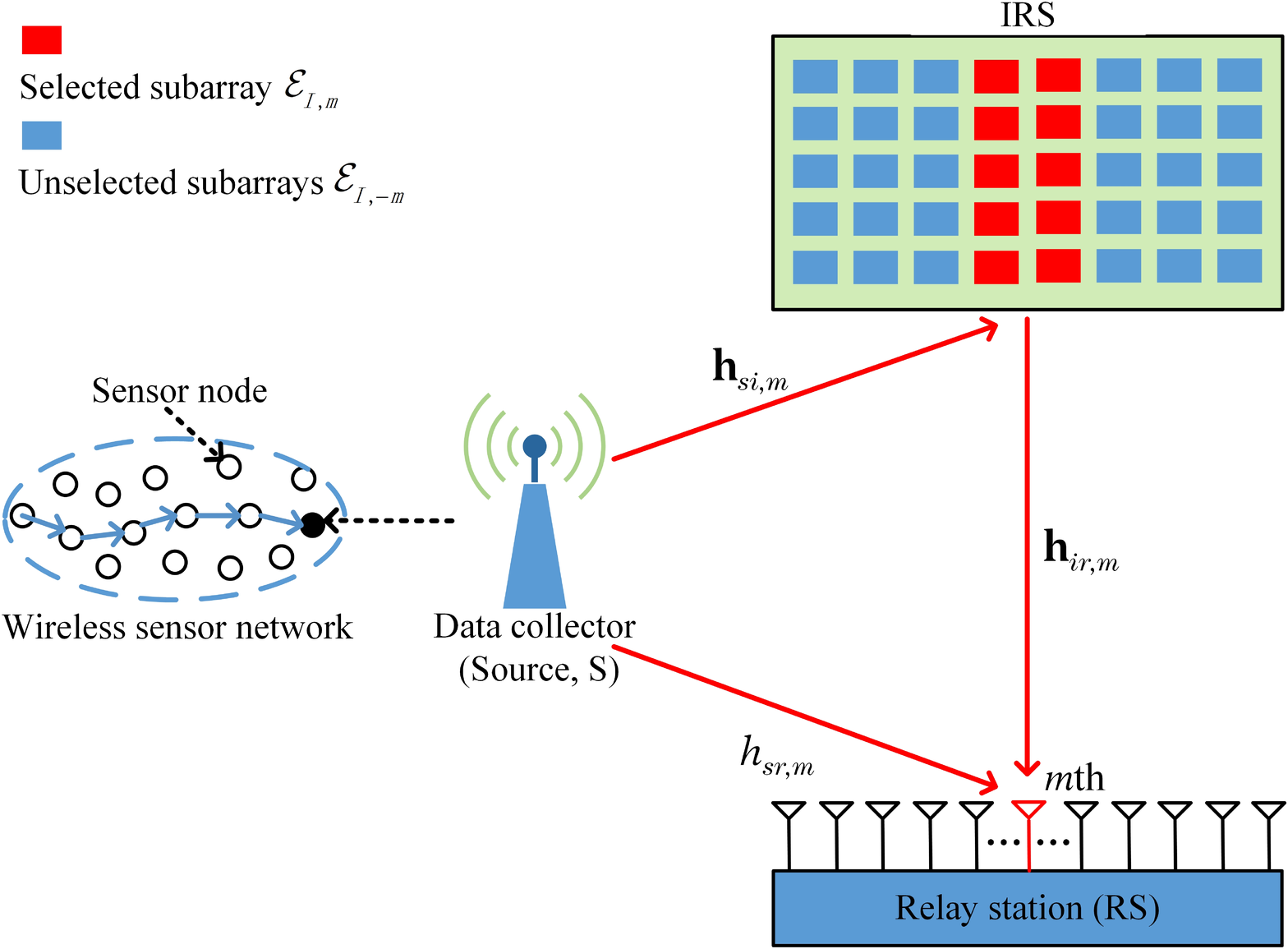}\\
\caption{System model for IRSES-based Max-RP plus MRC method in the first time slot.}\label{2System-model-for-method-C.eps}
\end{figure}
\begin{figure}[h]
\centering
\includegraphics[width=0.48\textwidth,height=0.09\textheight]{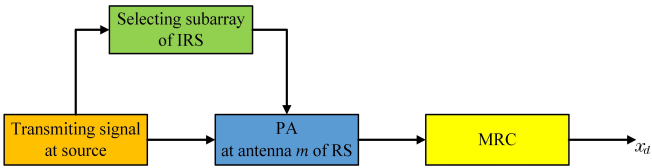}\\
\caption{Block diagram for IRSES-based Max-RP plus MRC method.}\label{3block-diagram-for-method-C.eps}
\end{figure}

A subarray denoted as ${{\cal E}_{I,m}}$ is randomly chosen to reflect the signal from S to antenna $m$ of RS, ${{\cal E}_{I,m}}$ = $\left\{ {{1_m},{2_m}, \cdots ,{K_m}} \right\}$. Let ${\cal E}_I$= ${\cal E}_{I,1} \cup {\cal E}_{I,2} \cdots  \cup {\cal E}_{I,M}$ be the subarray set. ${\cal E}_{I,1},{\cal E}_{I,2}, \cdots ,{\cal E}_{I,M}$ is a partition of set ${\cal E}_I$, and ${\cal E}_{I,m} \cap {\cal E}_{I,n}$= $\Phi$, where $m,n$ $\in \left\{ {1,2, \cdots ,M} \right\}$. The remaining subsets are denoted as ${\cal E}_{I,-m}$, and ${\cal E}_I$= ${\cal E}_{I,m}$ $\cup$ ${\cal E}_{I,-m}$. The received signal of antenna $m$ at RS can be written by
\begin{align}\label{emph{y}_{r,m}}
&y_{r,m}=\sqrt {P_s}(h_{sr,m}+\textbf{h}_{ir,m}\boldsymbol{\Theta}_{1,m}\textbf{h}_{si,m})s+\nonumber\\
&~~~~~~~~~~~~~~~~~~~\sqrt {P_s} \textbf{h}_{ir,-m}\boldsymbol{\Theta}_{1,-m}\textbf{h}_{si,-m}s+n_{r,m},
\end{align}
where $h_{sr,m}\in {\mathbb{C}^{1 \times 1}}$ is the $m$th element of $\textbf{h}_{sr}$,
$\textbf{h}_{ir,m}\in {\mathbb{C}^{1 \times K}}$ denotes the channel from the selected subset ${\cal E}_{I,m}$ to antenna $m$ at RS, and includes $K$ elements which are determined by ${\cal E}_{I,m}$ in the $m$th row of $\textbf{H}_{ir}$. $\textbf{h}_{si,m}\in {\mathbb{C}^{K \times 1}}$ represents the channel from S to ${\cal E}_{I,m}$ and also consists of $K$ elements decided by ${\cal E}_{I,m}$ in $K$ rows of $\textbf{h}_{si}$.
The diagonal reflection-coefficient matrix of elements in ${\cal E}_{I,m}$ is denoted as $\boldsymbol{\Theta}_{1,m}= \text{diag}({e^{j\theta _{1,m}(1)}}, \cdots ,{e^{j\theta _{1,m}(K)}})$. $\textbf{h}_{ir,-m}\in {\mathbb{C}^{1 \times (N-K)}}$ is the channel from the unselected subset ${\cal E}_{I,-m}$ to antenna $m$ at RS, $\textbf{h}_{si,-m}\in {\mathbb{C}^{(N-K) \times 1}}$ is the channel from S to ${\cal E}_{I,-m}$. $\boldsymbol{\Theta}_{1,-m}$ is the diagonal reflection-coefficient matrix of elements in ${\cal E}_{I,-m}$, which can be expressed as $\boldsymbol{\Theta}_{1,-m}= \text{diag}({e^{j\theta _{1,-m}(1)}}, \cdots ,{e^{j\theta _{1,-m}(N-K)}})$. $n_{r,m}$ is the AWGN with the distribution of $n_{r,m} \sim \mathcal{CN}\left( 0,{\sigma_{r,m }^2} \right)$.
While according to law of large numbers, the signals reflected by the unselected subset ${\cal E}_{I,-m}$ overlap to zero at antenna $m$. (\ref{emph{y}_{r,m}}) is simplified and expanded as follows
\begin{align}
y_{r,m}&=\sqrt {P_s}(h_{sr,m}+\textbf{h}_{ir,m}\boldsymbol{\Theta}_{1,m}\textbf{h}_{si,m})s+n_{r,m}\nonumber\\
&=\sqrt {P_s}\left(  h_{sr,m}+ \sum\limits_{i = 1}^K \textbf{h}_{ir,m}(i)\boldsymbol{\Theta}_{1,m}(i,i)\textbf{h}_{si,m} (i)\right )s+\nonumber\\
&~~~~~~~~n_{r,m},
\end{align}
where $\textbf{h}_{ir,m}(i)$ and $\textbf{h}_{si,m} (i)$ are the $i$th element of $\textbf{h}_{ir,m}$ and $\textbf{h}_{si,m}$, respectively. $\boldsymbol{\Theta}_{1,m}(i,i)$ is the $i$th element in diagonal of $\boldsymbol{\Theta}_{1,m}$.
The phases of $K$ elements in ${\cal E}_{I,m}$ are adjusted to make the phases of all reflected signals and the direct signal from S aligned at antenna $m$ of RS. So as to maximize the power of the signal received by antenna $m$ for obtaining better rate performance.
Thus, the corresponding phase shift of the $i$th element is calculated as
\begin{equation}
\theta _{1,m}(i)=\text{arg} (h_{sr,m})-\text{arg}(\textbf{h}_{ir,m}(i))-\text{arg}(\textbf{h}_{si,m}(i)).
\end{equation}
Finally, MRC is adopted to combine all received signals at RS, the received signal can be expressed as follows
\begin{equation}
y_r=[u_{r,1}, u_{r,2} ,\cdots,u_{r,M}][y_{r,1}, y_{r,2} ,\cdots,y_{r,M}]^T,
\end{equation}
where
\begin{equation}
u_{r,m}=\frac{(h_{sr,m}+\textbf{h}_{ir,m}\boldsymbol{\Theta}_{1,m}\textbf{h}_{si,m} )^H}{\|h_{sr,m}+\textbf{h}_{ir,m}\boldsymbol{\Theta}_{1,m}\textbf{h}_{si,m}\|}.
\end{equation}
The achievable rate at RS can be represented as
\begin{equation}
R_r=\log_2\left(1+\frac{\sum_{m=1}^M\|h_{sr,m}+\textbf{h}_{ir,m}\boldsymbol{\Theta}_{1,m}\textbf{h}_{si,m}\|^4P_s }{\sum_{m=1}^M\|h_{sr,m}+\textbf{h}_{ir,m}\boldsymbol{\Theta}_{1,m}\textbf{h}_{si,m}\|^2\sigma _{r,m}^2}\right).
\end{equation}

Further, taking the minimum of $R_r$ and $R_d$ to obtain achievable rate $R_s$. The complexity of the proposed IRSES-based Max-RP plus MRC method is given by
\begin{equation}
\mathcal{O}\{15MK+8M+10K+L_5(18MN+2M+3N)\},
\end{equation}
where $L_5$ is the AI number in the second time slot. It can be seen that the highest order of complexity is $MN$ FLOPs, which is much lower than the above two methods. Especially in massive IRS scenario, the gap among the complexity of the proposed three methods is more obvious.

\section{Simulation And Numerical Results}
In this section, numerical simulations are performed to evaluate and compare the rate performance between an IRS-aided multi-antenna DF relay network and that with single-antenna RS in \cite{2020A}.
Additionally, it is assumed that IRS and RS are deployed with the same abscissas,
the positions of S, RS, IRS, and D are given as (0,0), (50m,0), (50m,10m) and (100m,0), respectively. The amplitude attenuation of the received signal is $(d)^{-(\frac{\alpha}{2})}$ caused by path loss, where $d$ is the distance between transmitter and receiver, and $\alpha$ is the path-loss exponent. The antenna gains at S, RS and D are  5 dBi, 5dBi,  and 2 dBi, respectively.
System parameters are set as follows: $P_s$ = $P_r$ = 10dBW, $\alpha$ = 2.4, and $\sigma_{d}^2$ = $\sigma_{r}^2 = \sigma_{w}^2$. SNR is defined as $(P_s+P_r)/\sigma_{w}^2 $.

\begin{figure}[h]\label{IterCur}
\centering
\includegraphics[width=0.48\textwidth,height=0.27\textheight]{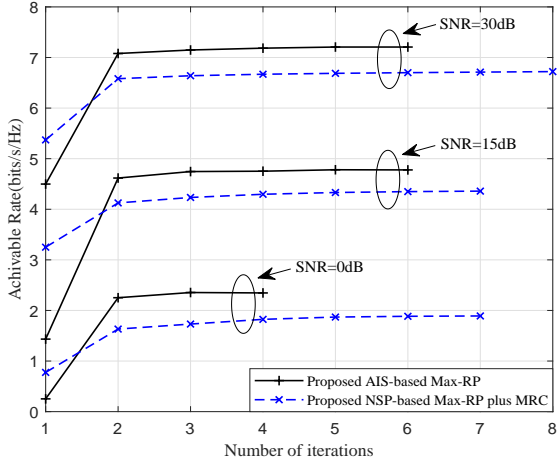}\\
\caption{Convergence of proposed methods with $M$ = 50 and $N$ = 50.}\label{4iteration-convergence.eps}
\end{figure}

\begin{figure}[h]
\centering
\includegraphics[width=0.48\textwidth,height=0.27\textheight]{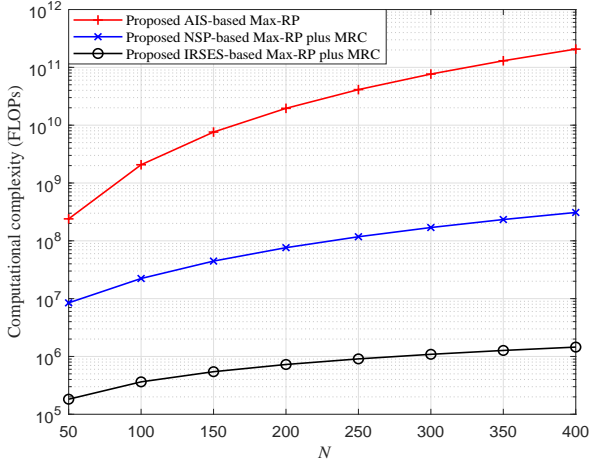}\\
\caption{Computational complexity versus $N$ with $M$ = 50.}\label{5complexity.eps}
\end{figure}

\begin{figure}[h]
\centering
\includegraphics[width=0.48\textwidth,height=0.27\textheight]{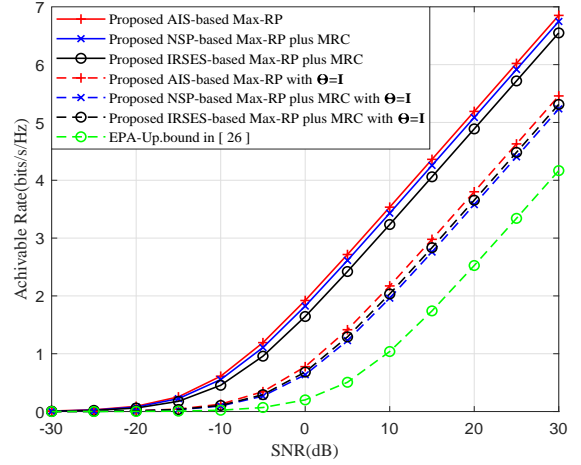}\\
\caption{Achievable rate versus SNR with $M$ = 16 and $N$ = 160.}\label{6R-VS-SNR-M16-N160.eps}
\end{figure}

\begin{figure}[h]
\centering
\includegraphics[width=0.48\textwidth,height=0.27\textheight]{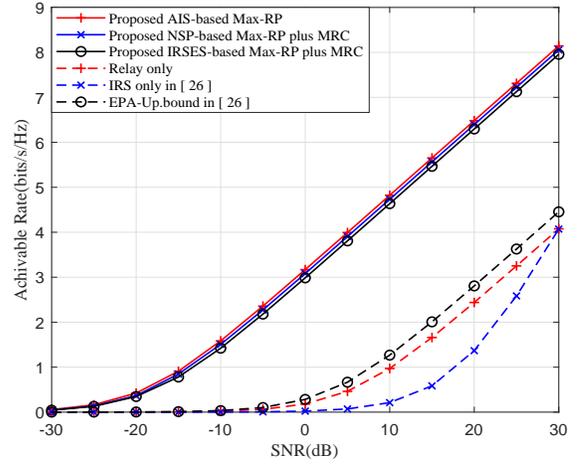}\\
\caption{Achievable rate versus SNR with $M$ = 50 and $N$ = 200.}\label{7R-VS-SNR-M50-N200.eps}
\end{figure}

\begin{figure}[h]
\centering
\includegraphics[width=0.48\textwidth,height=0.27\textheight]{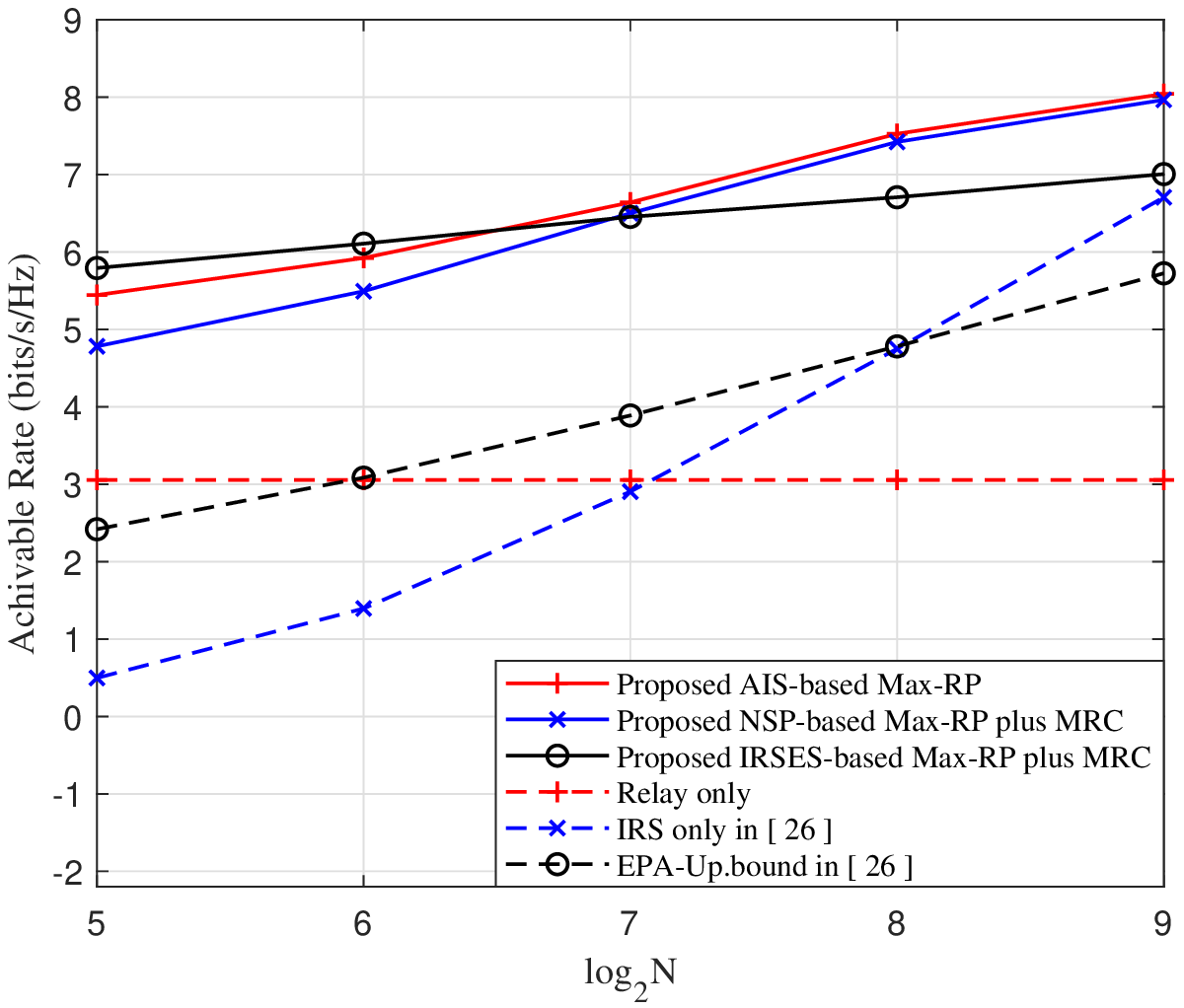}\\
\caption{Achievable rate versus $N$ with $M$ = 16 and SNR = 30\text{dB}.}\label{8R-VS-N-M16-SNR30.eps}
\end{figure}

\begin{figure}[h]
\centering
\includegraphics[width=0.48\textwidth,height=0.27\textheight]{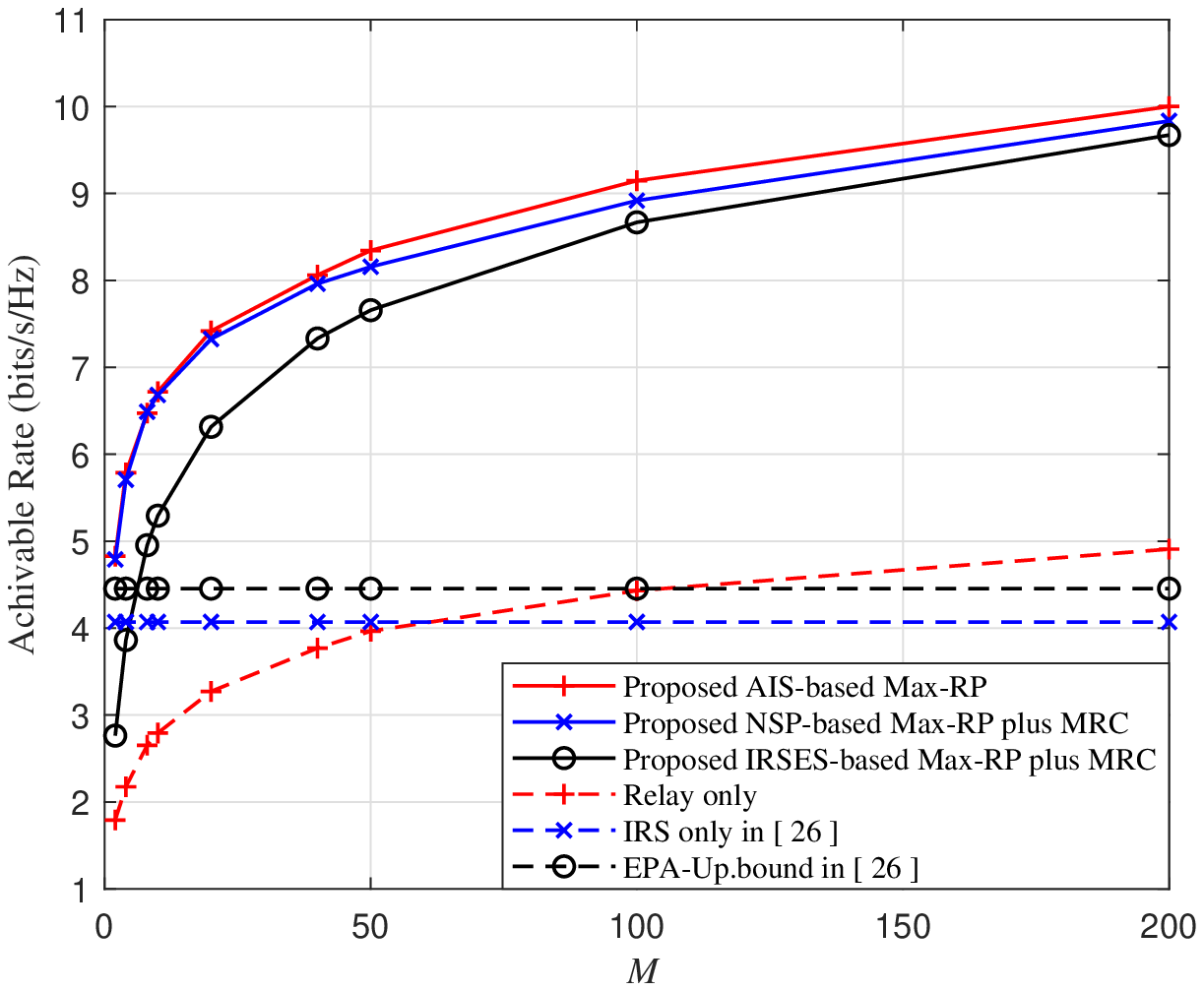}\\
\caption{Achievable rate versus $M$ with $N$ = 200 and SNR = 30\text{dB}.}\label{9R-VS-M-N200-SNR30.eps}
\end{figure}

\begin{figure}[h]
\centering      
\subfigure[]{
\begin{minipage}{8.6cm}
\centering
\includegraphics[width = 7.2 cm, height = 4.80cm]{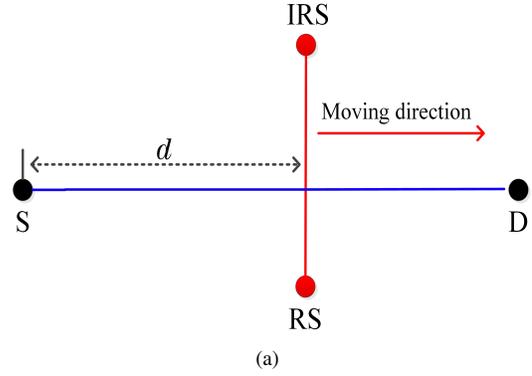}
\end{minipage}
}
\subfigure[]{
\begin{minipage}{8.6cm}
\centering      
\includegraphics[width = 8.72 cm, height = 6.55cm]{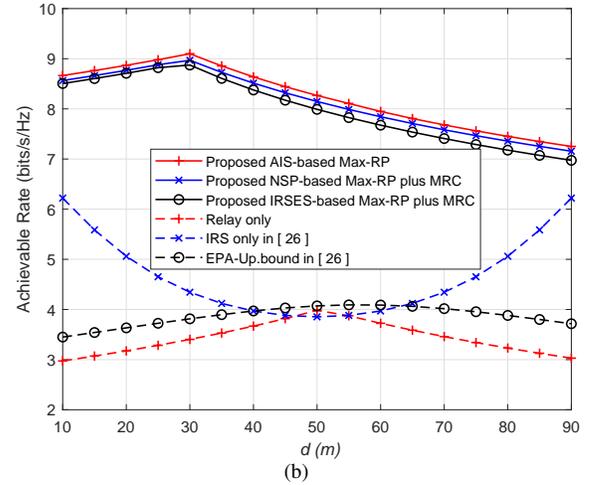}
\end{minipage}
}
\caption{(a) How to move IRS and RS. (b) Achievable rate versus distance with $M$ = 50, $N$ = 200 and SNR = 30\text{dB}.}
\end{figure}

Fig.~4 shows the achievable rate versus the number of iterations with $M$ = 50 and $N$ = 50 in the first time slot. It is obvious that the proposed AIS-based Max-RP and NSP-based Max-RP plus MRC methods require about only three iterations to achieve the rate ceil.

Fig.~5 plots the curves of computational complexity versus $N$ with $M$ = 50. It demonstrates that the computational complexities of the proposed three methods,  AIS-based Max-RP, NSP-based Max-RP plus MRC, and IRSES-based Max-RP plus MRC, increase as $N$ increases. Clearly, the first and third methods have the highest and lowest computational complexities, respectively. The second one is  in between them.

Fig.~6 shows the curves of achievable rate versus SNR with $M$ = 16 and $N$ = 160. It is clear that the rate performance of the proposed three methods with fixed IRS phases (i.e., ${\boldsymbol\Theta}$ = $\textbf{I}$) to eliminate the impact of IRS elements on the performance is more than 5.24 bits/s/Hz, which is 25.6\% higher than that of the existing system with an IRS and a single-antenna relay in \cite{2020A}. Compared to the above methods with ${\boldsymbol\Theta}$ = $\textbf{I}$, the performance of the proposed three methods with joint RS active beamforming and IRS passive beamforming can approximately harvest up to 25.1\% rate gain. Therefore, it is verified that the rate gain results from the increased number of antennas at RS and from the joint RS active beamforming and IRS passive beamforming.

Fig.~7 illustrates the curves of achievable rate versus SNR with $M$ = 50 and $N$ = 200. It can be seen that  the  proposed methods make a significant performance improvement  over that with single-antenna RS in \cite{2020A}.  For example, when  SNR equals 30dB, the proposed worst method, IRSES-based Max-RP plus MRC, can harvest up to 78.6\% rate gain over that method in \cite{2020A}. The best method AIS-based Max-RP approximately has a 80.8\% rate gain over
that in \cite{2020A}. This shows that as SNR increases, significant rate gains are achieved for the proposed network with IRS plus multi-antenna RS.

Fig.~8 demonstrates the achievable rate versus the number $N$ of reflecting elements at IRS. It is observed that the  proposed three methods still outperform the method with single-antenna RS in \cite{2020A}. For small-scale and medium-scale IRS, the proposed three methods have the following increasing order on rate as follows: NSP-based Max-RP plus MRC, AIS-based Max-RP and IRSES-based Max-RP plus MRC. As the number of elements at IRS goes to  a large-scale, their order are becomes as follows:  IRSES-based Max-RP plus MRC, NSP-based Max-RP plus MRC and AIS-based Max-RP.

Fig.~9 shows the curves of achievable rate versus $M$ with $N$ = 200 and SNR = 30dB. It can be seen that as the number $M$ of antennas at RS increases, the rate performance increases. The proposed three methods: AIS-based Max-RP, NSP-based Max-RP plus MRC and IRSES-based Max-RP plus MRC have harvested significant rate performance gains over that in \cite{2020A}, IRS-only-aided network and relay-only-aided network.

In order to observe the effect of positions of IRS and RS on rate performance, Fig.~10~(a) shows how to move both IRS and RS, where IRS and RS move toward D together along the direction  parallel to the line segment SD with $d$ denoting the horizontal distance from S to IRS and RS .
Fig.~10~(b) plots the corresponding curves of achievable rate versus $d$ with $M$ = 50, $N$ = 200 and SNR = 30\text{dB}. When $d$ = 30m, the proposed three methods can achieve their largest rate peak. While $d$ $>$ 30m, the rate performance  degrades gradually. Regardless of $d$, the proposed three methods: AIS-based Max-RP, NSP-based Max-RP plus MRC and IRSES-based Max-RP plus MRC still perform much better than that in \cite{2020A}, IRS-only-aided network and relay-only-aided network  in terms of rate performance.

\section{Conclusions}

In this paper, we have made an investigation on an IRS-aided DF relay network with multi-antenna at RS. In order to improve the rate performance, three high-performance schemes, namely AIS-based Max-RP, NSP-based Max-RP plus MRC and IRSES-based Max-RP plus MRC, were proposed. The third method is very attractive due to its low-complexity and excellent rate performance for small and medium -scale IRSs. Simulation results show that the proposed three methods can approximately harvest up to 85\% rate gain over existing network with single-antenna RS in almost all SNR regions. Thus, an IRS-aided  multi-antenna relay network will provide an enhanced network performance and extended network coverage for the future mobile communications, WSNs, and internet of things.

\ifCLASSOPTIONcaptionsoff
  \newpage
\fi

\bibliographystyle{IEEEtran}

\bibliography{IEEEfull,reference}

\end{document}